\documentclass[lettersize,journal]{IEEEtran}
\usepackage{amsmath,amsfonts}
\usepackage{algorithmic}
\usepackage{algorithm}
\usepackage{titlesec}
\usepackage{array}
\usepackage[caption=false,font=normalsize,labelfont=sf,textfont=sf]{subfig}
\usepackage{textcomp}
\usepackage{stfloats}
\usepackage{url}
\usepackage{multirow}
\usepackage{color}
\usepackage{verbatim}
\usepackage{graphicx}
\usepackage{cite}
\definecolor{r}{rgb}{1, 0, 0}
\definecolor{b}{rgb}{0, 0, 1}

\begin{document}

\title{\fontsize{21pt}{24pt}\selectfont Generative AI for Game Theory-based Mobile Networking}

\author{
    Long~He,
    Geng~Sun, Dusit~Niyato,~\IEEEmembership{Fellow,~IEEE}, 
    Hongyang~Du,
    Fang~Mei,
    Jiawen~Kang,
    Mérouane~Debbah,~\IEEEmembership{Fellow,~IEEE},
    and Zhu~Han,~\IEEEmembership{Fellow,~IEEE}
    \thanks{This research is supported by the National Natural Science Foundation of China (62272194, 62471200), the Science and Technology Development Plan Project of Jilin Province (20220101101JC), the National Research Foundation, Singapore, and Infocomm Media Development Authority under its Future Communications Research \& Development Programme, Defence Science Organisation (DSO) National Laboratories under the AI Singapore Programme (FCP-NTU-RG-2022-010 and FCP-ASTAR-TG-2022-003), Singapore Ministry of Education (MOE) Tier 1 (RG87/22), and the NTU Centre for Computational Technologies in Finance (NTU-CCTF). (\textit{Corresponding authors: Geng Sun and Fang Mei.)}}
    \IEEEcompsocitemizethanks 
    {\IEEEcompsocthanksitem Long He and Fang Mei is with the College of Computer Science and Technology, Jilin University, Changchun 130012, China (e-mail: helong0517@foxmail.com, meifang@jlu.edu.cn).    	
    \IEEEcompsocthanksitem Geng Sun is with the College of Computer Science and Technology, Jilin University, Changchun 130012, China, and also with the College of Computing and Data Science, Nanyang Technological University, Singapore 639798 (e-mail: sungeng@jlu.edu.cn).    
    \IEEEcompsocthanksitem Dusit Niyato is with the College of Computing and Data Science, Nanyang Technological University, Singapore 639798 (e-mail: dniyato@ntu.edu.sg).  
    \IEEEcompsocthanksitem Hongyang Du is with the Department of Electrical and Electronic Engineering, University of Hong Kong, Pok Fu Lam, Hong Kong (e-mail: duhy@eee.hku.hk). 
    \IEEEcompsocthanksitem Jiawen Kang is with the School of Automation, Guangdong University of Technology, Guangzhou, China (e-mail: kavinkang@gdut.edu.cn).    
    \IEEEcompsocthanksitem Mérouane Debbah is with Khalifa University of Science and Technology, P O Box 127788, Abu Dhabi, UAE (e-mail: merouane.debbah@ku.ac.ae).
    \IEEEcompsocthanksitem Zhu Han is with the University of Houston, Houston TX 77004, USA, and also with the Department of Computer Science and Engineering, Kyung Hee University, Seoul 446701, South Korea (e-mail: hanzhu22@gmail.com).    
    }
    }



\maketitle

\begin{abstract}

\par With the continuous advancement of network technology, various emerging complex networking optimization problems have created a wide range of applications utilizing game theory. However, since game theory is a mathematical framework, game theory-based solutions often rely heavily on the experience and knowledge of human experts. Recently, the remarkable advantages exhibited by generative artificial intelligence (GAI) have gained widespread attention. In this work, we propose a novel GAI-enabled game theory solution that combines the powerful reasoning and generation capabilities of GAI to the design and optimization of mobile networking. Specifically, we first outline the game theory and key technologies of GAI, and explore the advantages of combining GAI with game theory. Then, we review the contributions and limitations of existing research and demonstrate the potential application values of GAI applied to game theory in mobile networking. Subsequently, we develop a large language model (LLM)-enabled game theory framework to realize this combination, and demonstrate the effectiveness of the proposed framework through a case study in secured UAV networks. Finally, we provide several directions for future extensions.

\end{abstract}


%
%

\section{Introduction}

\par Game theory studies mathematical models of strategic interactions among interdependent rational participants (referred to as players), who make decisions based on their self-interests while their actions influence one another. Given its highly abstract representation of real-life situations, game theory has emerged as a standard analytical tool in various fields to model and comprehend complex systems involving multiple interacting agents. In particular, with the development of wireless technologies, game theory plays an increasingly important role in mobile networking optimization, such as network configuration
and management, as well as network security and privacy protection~\cite{Charilas2010}. For example, a game theory-based method~\cite{lai2022online} was proposed to optimize user allocation and transmission power to improve the performance of mobile edge computing networks. Employing game theory for mobile networking optimization requires both constructing a suitable abstract game model (i.e., a binding from the cases to game-theoretic models) and solving the game (i.e., running an algorithm on the constructed model to extract the Nash equilibrium). However, addressing the aforementioned issues can be challenging since it traditionally requires a deep understanding of game theory and solution algorithms, particularly for the newcomers or those with interdisciplinary backgrounds. 

\par Fortunately, generative artificial intelligence (GAI) techniques, such as large language models (LLMs) combined with retrieval-augmented generation (RAG), offer a promising solution to address the aforementioned challenges. First, LLMs, such as ChatGPT-4, have demonstrated remarkable capabilities in natural language understanding and the generation of diverse content, with the potential to enable intelligent applications of game theory. Specifically, by offering natural language descriptions of network optimization problems, the LLMs have the potential to analyze these problems by using appropriate game theory models and generate effective solution algorithms, simplifying the use of game theory. Moreover, RAG technology can maintain a large-scale knowledge base that stores relevant knowledge on network optimization, game theory, and algorithms. By providing the LLMs with extensive specialized information, they can enhance the accuracy of content generation while overcoming the high dependency on specialized knowledge required for applying game theory.

\par Motivated by these, this work attempts to provide a forward-looking study to explore the use of LLMs and RAG technology to achieve the intelligent formulation and solution of game theory models by providing natural language descriptions of mobile networking optimization. To this end, we propose an LLM-enabled game theory framework, which implements the LLM module using the ChatGPT-4 model to understand human language, and enhances the generation ability of LLMs by building the RAG module using Langchain. To the best of our knowledge, this is the first work to systematically demonstrate the use of GAI technology to achieve the intelligence of game theory in mobile networking optimization. The contributions of this work are summarized as follows:

\begin{enumerate}
    \item We overview the game theory and introduce different types of GAI models. Building on this, we provide comprehensive insights into the integration of GAI and game theory.
    \item We explore a few seminal applications of LLMs and game theory in networking from different perspectives, providing the guideline on how to integrated these important technologies to solve practical problems.
    \item We propose an LLM-enabled game theory framework for networking optimization. Simulation results based on a real-world networking optimization case study validate the effectiveness of the proposed framework.
\end{enumerate}

%
%

\section{Overview of Game Theory and Generative AI}

\par In this section, we first introduces the basics of game theory and several typical AI technologies that forming the foundation of GAI. Following that, we generally explain the potential support of GAI for game theory.

\begin{figure*}[h] 
	\centering
	\setlength{\abovecaptionskip}{2pt}%
	\setlength{\belowcaptionskip}{2pt}%
	\includegraphics[width =6.7in]{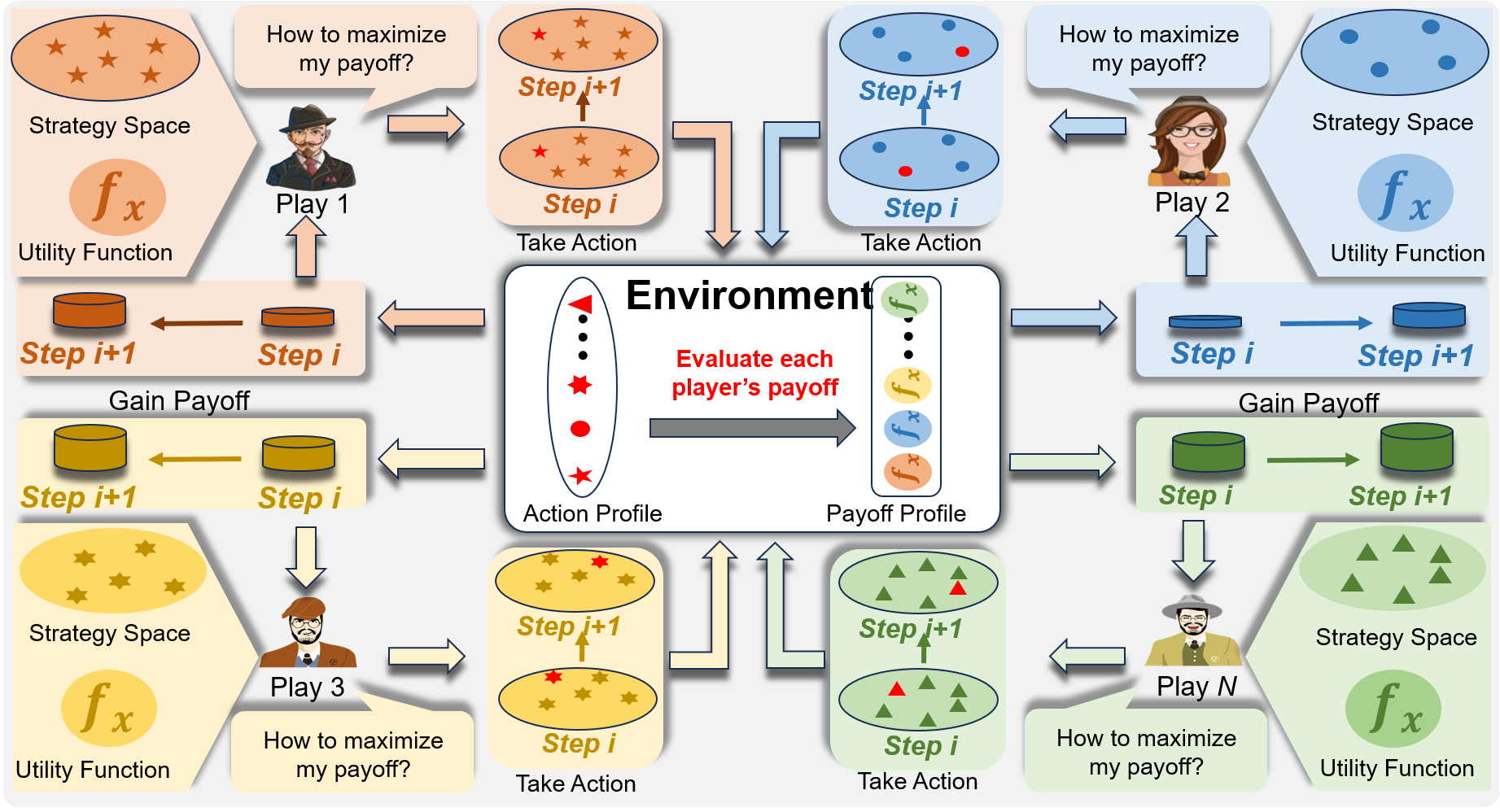}
	\caption{The non-cooperative game framework contains multiple players and an environment in which players interact. Each player has a strategy space that represents the available actions and a utility function that evaluates the player's payoff from taking a certain action. Additionally, the objective of each player is to take actions that maximize their own payoff. The environment represents the medium through which players interact, that matches an action profile to a payoff profile. The flow of player interaction is as follows: in the first interaction (Step 1), each player takes an action from their strategy space and receives the corresponding payoff. In subsequent interactions (Step i), each player updates their action based on the observed strategies of other players to improve their own payoff. When the actions of all players no longer change, a Nash equilibrium is obtained.}
	\label{fig_gameModel3}
	\vspace{-1em}
\end{figure*}

%
%

\subsection{Game Theory}

\par \textit{1) What is game theory about?} Game theory is a mathematical model and theoretical tool that studies the interaction of rational decision-makers (called players) in the presence of conflicts of interest. For example, the time required for a car driver to get home typically depends not only on the route they choose but also on the decisions made by other drivers. For such scenarios, game theory provides a framework for analyzing the strategic behavior of rational decision-makers and predicting the outcomes of their decisions. In game theory, non-cooperative game theory (NGT) is one of the most elemental and important branches, which has been widely applied in various fields. Specifically, it typically implies a competitive nature where players are not allowed to engage in enforced cooperation. Therefore, in this brief exposition, this paper focuses on a review of NGT.

\par \textit{2) How to present a game?} The representation of game theory typically involves mapping a real-world game problem or scenario into a mathematical model. For NGT, as shown in Fig.~\ref{fig_gameModel3}, the strategic form is the most basic form, which typically consists of three fundamental elements that are players, strategies and payoffs.

\begin{itemize}
    \item \textbf{Players:} Players are the participants or decision-makers involved in the game. A player can be a person, a machine, or more generally, any decision-making entity. Each player has a distinct set of preferences and objectives, and their choices or actions impact the outcomes of the game.
    \item \textbf{Strategies:} Strategies represent the available choices or actions that players can take in the game. A strategy of the player is their plan of action, which determines how they will act or respond to the actions of other players. 
    \item \textbf{Payoffs:} Payoffs in game theory represent the objectives or utilities that players receive based on the strategies they choose and the actions taken by other players. The payoffs determine the desirability of different outcomes for each player, influencing their decisions and strategies.
\end{itemize}

\par \textit{3) How to analyze the outcome of a game?} For NGT, the Nash equilibrium is a fundamental solution concept for predicting the outcome of the game. Specifically, it represents a strategy profile in which no player can improve their own payoff by changing their strategy while keeping the strategies of the other players unchanged. In other words, Nash equilibrium is a stable state where the players lack the incentive to actively change their strategies.

%
%

\subsection{Generative AI Models}

\par GAI is a collection of various AI technologies and models that can learn patterns and regularities from data and then generate new data. The foundation of GAI mainly includes the following key technologies.

\par \textbf{LLMs:} LLMs are AI models based on Transformers, which can understand and generate human language through training of large-scale text data. Due to their versatility and robust generalization capabilities, LLMs can undertake diverse tasks, including text classification, question answering systems, and text generation, etc.

\par \textbf{RAG:} RAG is an advanced natural language processing technique that integrates information retrieval and generative models to enhance the quality and relevance of generated text. Furthermore, when combined with RAG, LLMs can better understand user queries and generate text results that more accurately fulfill user needs by accessing external databases.

\par \textbf{Generative Adversarial Networks (GANs):} GANs are a type of neural network architecture with custom adversarial learning objectives, which consist of a generator and a discriminator. The goal of GANs is to train a generator network to generate samples that are similar to real data, and to train a discriminator network to distinguish between samples generated by the generator and real samples. Therefore, GANs have high performance in generating high-quality multimodal content, which has led to a large number of applications such as 3D object generation, and image processing.

\par \textbf{Generative Diffusion Models (GDMs):} Inspired by non-equilibrium thermodynamics, GDMs comprise of two interconnected processes, i.e., a predefined forward process that transforms the data distribution into a more simplistic prior distribution (typically a Gaussian distribution), and a corresponding reverse process that employs a trained neural network to gradually reverse the effects of the forward process. Due to the ability to generate high-quality data and the advantage of modeling complex data distributions, GDMs have been employed in material design, time series forecasting, and text-to-image generation, etc.

%
%

\subsection{GAI for Game Theory}

\par Traditional AI technologies such as deep learning and reinforcement learning, have been applied in various successful cases in the context of game theory. For example, in~\cite{xu2023joint}, the authors formulated the task offloading problem in vehicular edge computing networks as an exact potential game and utilized a multi-agent distributed distributional deep deterministic policy gradient approach to attain the Nash equilibrium. In~\cite{abegaz2022blockchain}, the authors investigated the dynamic resource trading problem of multi-UAV-assisted industrial IoT networks, which is modeled as an extended stochastic game. Then, the authors proposed a multi-agent deep reinforcement learning algorithm to solve the formulated stochastic game. In~\cite{kang2024tiny}, a Stackelberg game was employed to model real-time UAV twin migration for the emerging UAV metaverse system. The authors designed a tiny multi-agent deep reinforcement learning algorithm to approximate the game equilibrium. However, traditional AI techniques applied in the context of game theory still encounter the following {\em challenges}:

\begin{itemize}
    \item \textbf{Need for Formal Formulation:} Traditional AI techniques typically require a formal formulation of a game-theoretic scenario, i.e., mathematical representations, including the definition of players, actions, and payoffs. However, this formalization process can be challenging, especially for complex and real-world games where the rules may not be explicitly defined or known.
    \item \textbf{Limited Information and Communication:} In certain games, players may lack complete information about the game state or the strategies of other players. Traditional AI techniques may struggle to handle such incomplete or asymmetric information settings, as they rely on the assumption of complete information in their learning and decision-making processes.
    \item \textbf{Computational Complexity:} Obtaining Nash equilibrium for large games can incur significant computational costs. The search space for the Nash equilibrium increases exponentially with the number of players and strategies~\cite{babichenko2014simple}. Traditional AI techniques may struggle to efficiently explore the strategy space and converge to a Nash equilibrium in such cases.
\end{itemize}

\par LLMs combined with RAG technology provide a promising solution for addressing the abovementioned challenges faced by traditional AI. In the following, we elaborate on the advantages of LLMs combined with RAG technology in game scenarios from three perspectives, focusing on game scenario recognition, game theoretic model formulation, and Nash equilibrium search.

\begin{itemize}
    \item \textbf{Game Scenario Recognition:} Game theory has been widely used in various fields to analyze complex real-life situations. Applying game theory to case studies involves initially identifying and extracting key game theory-related concepts from the context of these cases, including decision-makers, their strategies, and their payoffs. LLMs, trained on large amounts of textual data, possess the capability to comprehend natural language, thereby accurately identifying game theory-related concepts from natural language descriptions of real-life cases.
    \item \textbf{Game Theoretic Model Formulation:} Game theoretic model formulation involves mapping a game scenario to a mathematical model to facilitate theoretical analysis. Due to the diversity and complexity of game scenarios, it is challenging to formulate suitable mathematical models, requiring a broad and in-depth understanding of game theory and mathematical method. LLMs combined with RAG has powerful learning and generation capabilities, which can formulate accurate mathematical models that are suitable for gaming scenarios through extensive learning of knowledge in related fields.
    \item \textbf{Nash Equilibrium Search:} Obtaining the outcomes of game theoretic models, i.e., the Nash equilibrium, often requires searching through large strategy spaces. Therefore, designing high-performance algorithms is crucial for reducing computational complexity. The combination of LLMs and RAG can generate high-performance algorithms by learning existing methods to obtain Nash equilibrium more efficiently.
\end{itemize}

\par \textbf{Lesson Learned:} The exploration of LLMs combined with RAG in game theory  underscores the critical role of rich knowledge base and powerful reasoning abilities in simplifying the utilization of game theory. Specifically, traditional game theory typically demands researchers to possess extensive experience and a deep understanding of game theory to provide effective insights into various complex game scenarios. By engaging in self-learning from large-scale data, LLM-based game theory successfully encapsulates the aforementioned complex process, thereby effectively reducing the challenges faced in the applications of traditional game theory. For example, in the social sciences domain, the integration of LLMs with game theory provides a valuable tool for experimental research and social simulations~\cite{guo2023gpt}. In the field of economics, LLMs have the potential to revolutionize microeconomic analysis by simulating strategic interactions among economic agents, which can lead to a better understanding of market behaviors and policy impacts~\cite{duan2024gtbench}.

%
%

\section{LLMs-enabled Game Theory in Networking}

\par In this section, we first provide a comprehensive review of the existing literature on the integration of LLMs and game theory, and analyze the limitations of the current works. Subsequently, we present some applications of LLM-enabled game theory in networking and analyze the combination of LLMs and game theory for multi-agent network design.

%
%

\subsection{Overview of Survey Papers}

\par Several studies have explored the integration of LLMs and game theory. Next, we will briefly review them and discuss their relevance to this work. Moreover, Table \ref{Table 1} provides a summary. 

\begin{table*}
	\setlength{\abovecaptionskip}{5pt}%
	\setlength{\belowcaptionskip}{5pt}%
	\caption{Overview of survey papers on the integration of LLMs and game theory}
	\label{Table 1}
\begin{tabular}{c|c|c}
\hline
\textbf{Survey} & \textbf{Contributions}  & \textbf{Emphasis} \\  
\hline
 \cite{fan2023can} & \multicolumn{1}{l|}{\begin{tabular}{p{10cm}}Evaluate the degree of rationality of LLMs in establishing clear desires, refining beliefs about uncertainty, and taking optimal actions in the context of  game theory.\end{tabular}} &  \multirow{2}*{\begin{tabular}{p{5cm}} Focus on LLMs' reasoning capabilities. \end{tabular}}\\
\cline { 1 - 2 }
\cite{duan2024gtbench} & \multicolumn{1}{l|}{\begin{tabular}{p{10cm}}Analyze the reasoning capabilities of LLMs in competitive environments through game-theoretic tasks that require pure logic and strategic reasoning to compete with opponents.\end{tabular}} &  \\
\hline
\cite{guo2023gpt} & \multicolumn{1}{l|}{\begin{tabular}{p{10cm}}Explore the use of LLMs in strategic game experiments, specifically the ultimatum game and the prisoner’s dilemma.\end{tabular}} &  \begin{tabular}{p{5cm}} Focus on the application of LLMs in strategy game experiments. \end{tabular}\\ 
\hline
\cite{zou2023wireless} & \multicolumn{1}{l|}{\begin{tabular}{p{10cm}}Investigate multi-agent LLMs from a game theoretic perspective, where agents collaboratively solve tasks in competitive environments.\end{tabular}} & \multirow{2}*{\begin{tabular}{p{5cm}} Focus on LLMs' collaborative capabilities.\end{tabular}} \\ 
\cline { 1 - 2 }
\cite{wu2024shall} & \multicolumn{1}{l|}{\begin{tabular}{p{10cm}} Explore the potential for LLM agents to spontaneously establish collaborative relationships in three competitive game scenarios. \end{tabular}} &  \\
\hline
\cite{gemp2024states} & \multicolumn{1}{l|}{\begin{tabular}{p{10cm}} Present the integration of game-theoretic frameworks with LLMs, enhancing the strategic capabilities of LLMs in dialogue by modeling conversations as games.\end{tabular}} &  \begin{tabular}{p{5cm}} Focus on the strategic capabilities of LLMs in dialogue. \end{tabular}\\ 
\hline
\end{tabular}
\end{table*}

\par Some works explored the reasoning capabilities of LLMs. For example, Fan \emph{et al.}~\cite{fan2023can} evaluated the degree of rationality of LLMs in establishing clear desires, refining beliefs about uncertainty, and taking optimal actions in the context of game theory. However, the context of game theory is pre-established. Moreover, this work lacks an evaluation of more practical scenarios, such as network optimization. Duan \emph{et al.}~\cite{duan2024gtbench} analyzed the reasoning capabilities of LLMs in competitive environments through game-theoretic tasks that require pure logic and strategic reasoning to compete with opponents. However, this work utilized LLMs to generate actions for various game scenarios based on the predefined prompt templates, which can affect the outcomes due to the prompt sensitivity. Guo \emph{et al.}~\cite{guo2023gpt} explored the use of LLMs in strategic game experiments, especially for the ultimatum game and the prisoner's dilemma. However, these strategic game cases all have clear game theory models, and whether the LLMs can map real-world scenarios to these models remains to be explored.

\par There are also studies focused on evaluating the collaborative capabilities of LLMs. For example, Zou \emph{et al.}~\cite{zou2023wireless} investigated the multi-agent LLMs from a game theoretic perspective, where the agents collaboratively solve tasks in the competitive environments. However, the proposed framework does not consider the issue of hallucinations in LLMs, where the generated content may be nonsensical or factually incorrect. In particular, model hallucinations may occur when the generative agents fail to capture the rapidly changing dynamics of wireless networks, leading to inaccurate representations of the wireless environment. Wu \emph{et al.}~\cite{wu2024shall} explored the potential for LLM agents to spontaneously establish collaborative relationships in three competitive game scenarios. However, this work only explores the potential of LLMs without incorporating the latest RAG technology. It is worth exploring how utilizing RAG to store extensive knowledge could enhance the ability of LLMs to establish collaborative relationships.

\par Moreover, Gemp \emph{et al.}~\cite{gemp2024states} investigated the integration of game-theoretic frameworks with LLMs, enhancing the strategic capabilities of LLMs in dialogue by modeling conversations as games. However, whether the strategic reasoning capabilities of LLMs in language understanding can be effectively applied to game-theoretic scenarios to enhance the practical applications of game theory remains underexplored.

\par In general, the abovementioned studies have the following shortcomings compared to our work. First, the studies above mainly focus on exploring the reasoning capabilities of LLMs in the context of game theory. However, the potential advantages of applying LLMs to game theory scenarios have not been well studied. Second, the abovementioned studies are mainly based on classical game theory models with pre-defined game theory models. However, in various application fields of game theory such as the networking, it is challenging to map actual game scenarios in real world into appropriate game theory models. Given the potent natural language understanding and generation capabilities of LLMs, leveraging them to map game scenarios to game models is a great value topic, which is still lacking research. Finally, the abovementioned studies do not consider the combination of LLMs with RAG. However, RAG is crucial to improve the capabilities of LLMs. Therefore, the combination of LLMs and RAG needs further exploration. To this end, this paper proposes a framework that applies LLMs combined with RAG to game theory in networking scenarios to make up for the shortcomings of existing research.
\begin{figure*}[h] 
    \centering
    \setlength{\abovecaptionskip}{1pt}%
    \setlength{\belowcaptionskip}{1pt}%
    \includegraphics[width =7in]{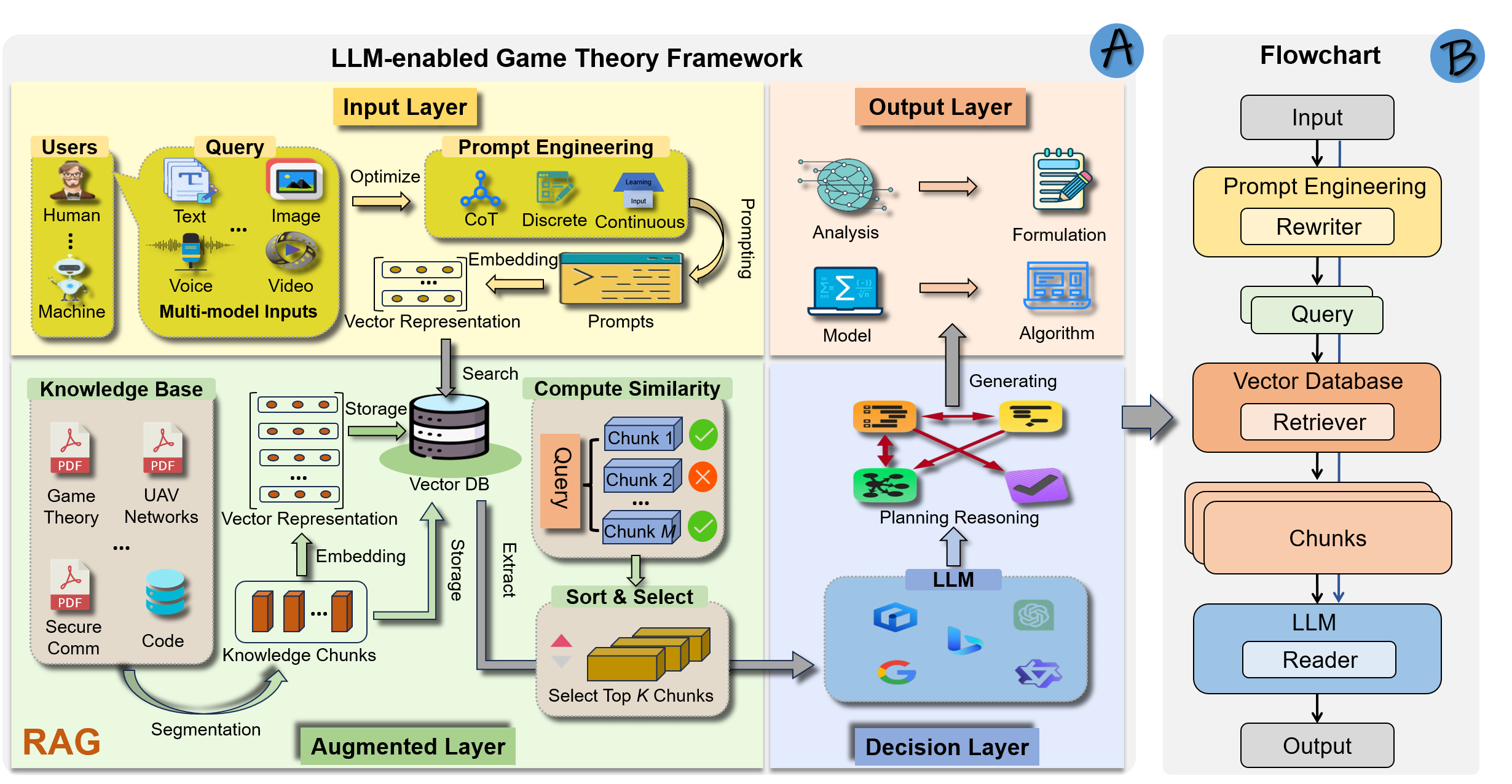}
    \caption{Part A shows the LLM-enabled game theory framework. The framework is based on a layered architecture consisting of an input layer, an augmented layer, a decision layer, and an output layer. The input layer captures user input queries. The augmented layer employs RAG technology to enhance user queries. The decision layer utilizes a pluggable LLM module to generate responses. The output layer returns the generated results to the user. Part B shows the flowchart of the proposed framework.}
    \label{fig_framework}
    \vspace{-1em}
\end{figure*}

%
%

\subsection{Potential Applications of LLMs-enabled Game Theory in Networking}

\par \textbf{Spectrum and Interference Management:} Spectrum and interference management are important issues in the field of wireless communication networks, involving the effective use of limited spectrum resources and power control to reduce wireless interference and improve network performance. Unlike traditional centralized management solutions with high computing load, game theory can be used to formulate distributed management solutions to effectively alleviate the computing load. However, traditional game theory approaches require frequent information interactions between the service users and central controller to obtain real-time complete information, which results in not only substantial time taken by users but also heavy computation and communication overhead. In this case, the LLMs-enabled game theory approach can use pre-trained models to approximate and learn strategic interaction behaviors in games to predict user decisions, which supplements the integrity of information to reduce the communication overhead. 

\par \textbf{Resource Allocation:} Resource allocation involves the effective allocation and management of limited network resources, such as communication resources, computing resources, and storage resources, to meet the needs of network users. The limited nature of network resources usually leads to competition among network users, while game theory provides an analytical framework to understand the decision-making and strategy selection process among users to obtain effective resource allocation solutions. However, traditional game theory approaches to evaluate user satisfaction usually rely on pre-defined models, which are inflexible and inaccurate in practical scenarios. LLMs-enabled game theory approaches can develop a more accurate model for evaluating the user satisfaction by combining historical information and user feedback, which can improve the effectiveness of decision-making.

\par \textbf{Network Security:} Security constitutes a critical aspect of network management, which involves various security controls and protocols to ensure the confidentiality of network resources. Game theory can be used to analyze the interaction between cyber attacks and defense strategies. By modeling the game between attackers and defenders, optimal defense strategies can be derived, thereby improving the network security performance. Traditional game theory approaches typically formulate optimal defense strategies based on the static network environments. However, in dynamic network environments, the strategies and behaviors of attackers may evolve over time, which is challenging for dealing with. In this case, LLM-based game theory can monitor the changing network security threats in order to adjust decision-making strategies to deal with the new attacks.

%
%

\subsection{Combining LLMs and Game Theory for Multi-Agent Network Design}

\par The convergence of LLMs, mobile networks, and multi-agent systems represents groundbreaking synergies that are leading to the emergence of multi-agent LLM network architectures, bringing huge potential for the design of future mobile networks~\cite{zou2023wireless}. This integration harnesses the power of collective intelligence and paves the way for self-managing networks. However, in a multi-agent LLM network, each LLM agent not only needs to pursue its individual objectives but also collaborate with other agents to achieve the collective goals of the entire network. This balance is crucial in designing efficient multi-agent LLM networks. In this context, game theory can be used to model and analyze the behavior of multi-agent LLM systems. This includes finding equilibria among agents, enabling them to serve individual objectives while effectively collaborating to achieve network-level goals. Furthermore, multi-agent reinforcement learning (RL) can further model the interactions among agents to learn optimal collaborative strategies and communication protocols among distributed LLM agents. In doing this, the communication costs among agents can be reduced.

%
%

\section{LLM-enabled Game Theory Framework Augmented by RAG for Mobile Network Optimization}
\begin{figure*}[h] 
	\centering
	\setlength{\abovecaptionskip}{2pt}%
	\setlength{\belowcaptionskip}{2pt}%
	\includegraphics[width =7in]{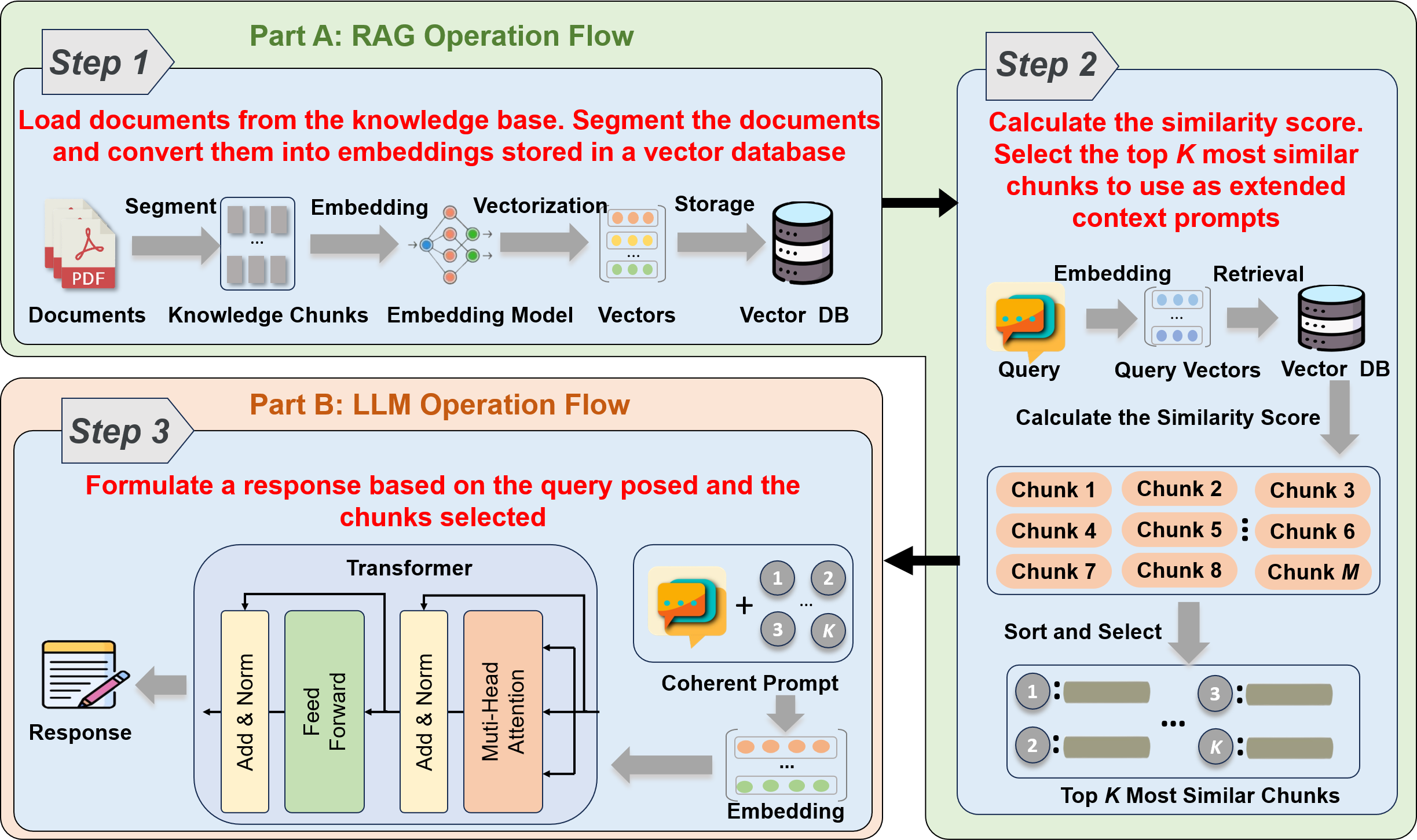}
	\caption{The operation flow of RAG models and LLMs. In Part A, documents are loaded from the knowledge base, segmented into chunks, encoded into vectors, and stored in the vector database. Then, RAG retrieves the $K$ most relevant chunks to the query of user based on semantic relevance from the vector database. In Part B, the original query and the retrieved chunks are inputted together into LLMs to generate the final response.}
	\label{fig-5}
	\vspace{-1em}
\end{figure*}

\par In the section, we first propose an LLM-enabled game theory framework for network optimization. Then, we conduct a case study on UAV secure communication optimization to demonstrate the effectiveness of the proposed framework.

%
%

\subsection{Motivations} Network optimization is crucial for enhancing the performance, efficiency, and user experience of mobile networks. Given the complexity of network optimization problems, game theory has been adopted and has played a significant role in analyzing various network optimization issues. Nevertheless, for network designers, employing game theory to analyze networking optimization problems pose challenges that typically necessitate a solid grasp of mathematical theory and a profound comprehension of game theory principles. 

\par Inspired by the outstanding capabilities of LLMs, we propose an LLM-enabled game theory framework that mainly consists of an LLM module and a RAG module. Specifically, the LLM module can automatically complete the mathematical formulation of network optimization problems, analyze the formulated optimization problems by using appropriate game theory models, and generate corresponding solutions through simple interactions with the user. Moreover, the RAG module contains a large-scale knowledge base that stores rich material data such as academic papers from IEEE Xplore, and it can improve the accuracy of LLM-generated responses by retrieving knowledge relevant to the user requests.

%
%

\subsection{Proposed Framework}

\par As shown in Fig. \ref{fig_framework}, our proposed framework follows a layered architecture, which consists of four layers that are the input layer, augmented layer, decision layer, and output layer.

\begin{itemize}
    \item \textbf{Input Layer:} The input layer receives requests from users, where multi-modal input allows the users to use different types of data as input, such as text, image, voice, and video. After receiving a request of the user, prompt engineering guides the system to generate the desired output by designing and constructing appropriate input prompts. Then, the input request of the user is transformed into a dense vector representation that the system can understand and process through embedding to facilitate downstream model processing. Furthermore, through embedding, the system can learn semantic similarities and correlations between inputs.
    \item \textbf{Augmented Layer:} The augmented layer is primarily implemented through the RAG technique, which is shown in detail in Part A of Fig. \ref{fig-5}. The RAG can improve the quality, accuracy, and relevance of the generated answers of the system by providing relevant information related to the user requests~\cite{asai2023self}. Specifically, The application of RAG consists of two phases, i.e., the database preparation phase and the knowledge retrieval phase. \textit{1) Database Preparation Phase}: RAG contains a large knowledge base, which stores sufficient external information, such as academic papers from IEEE Explore, technical documents, and algorithms from users. The information above is first divided into smaller knowledge chunks through semantic chunking techniques to facilitate embedded retrieval. Subsequently, a semantic embedding model is used to encode these knowledge chunks into vector representations, which are stored in an efficient vector database. \textit{2) Knowledge Retrieval Phase}: When the user inputs a query, RAG uses the same embedding model for constructing the vector database to convert the user query into a vector representation, which is then used to retrieve the vector database. Subsequently, RAG ranks the retrieved results based on vector similarity and returns the most relevant knowledge chunks. These retrieved knowledge chunks along with the user query form a prompt, which is sent to the decision-making layer as a foundational input for reasoning and decision-making by the generative model.
    \item \textbf{Decision Layer:} The decision layer adopts a plugin architecture, allowing it to select an appropriate LLM (such as New Bing, Bard, and GPT-4) to perform reasoning tasks according to the query. As shown in Part B of Fig. \ref{fig-5}, the LLMs generates responses based on the query and the relevant knowledge blocks retrieved from the augmentation layer. This approach enables the LLMs not only to leverage its internal language understanding and generation capabilities, but also to dynamically access external knowledge, resulting in more precise responses.
    \item \textbf{Output Layer:} The output layer presents the decision results given by the decision layer to the users.
\end{itemize}

%
%

\subsection{Case Study: UAV Secure Communication Optimization}
\begin{figure*}[h] 
	\centering
	\setlength{\abovecaptionskip}{2pt}%
	\setlength{\belowcaptionskip}{2pt}%
	\includegraphics[width =7in]{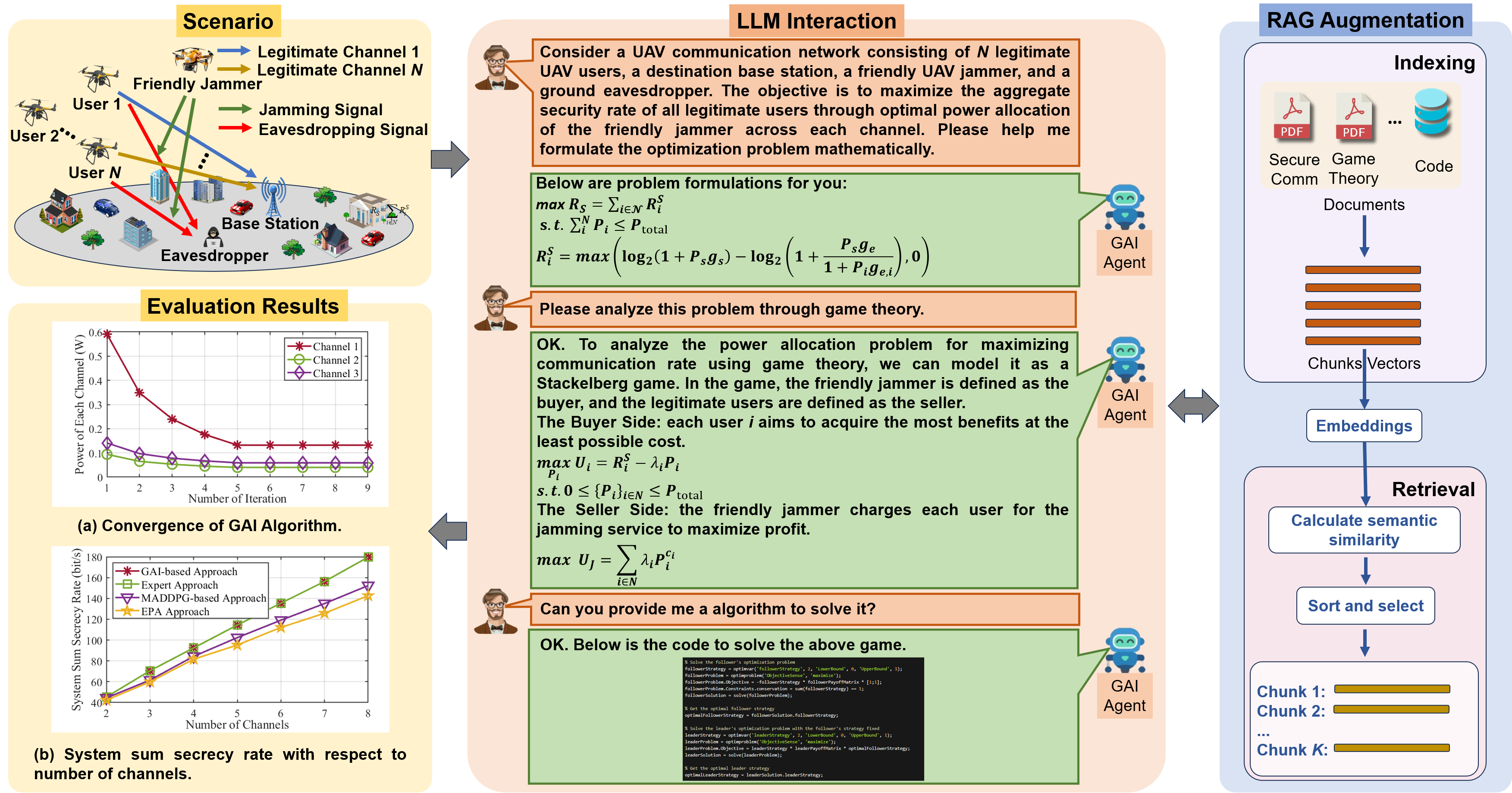}
	\caption{The experiment results of the UAV secure communication optimization case. The scenario module presents a system model of the considered scenario. The interaction module showcases the functionality of the proposed framework. The RAG augmentation module emphasizes the operational mechanism of RAG. The evaluation results module demonstrate the effectiveness of the proposed framework.}
	\label{fig_4}
	\vspace{-1em}
\end{figure*}

\par In the section, we use a UAV secure communication optimization case to demonstrate the proposed framework.

\par \textbf{Scenario Description:} As shown in Fig. \ref{fig_4}, consider a UAV communication network consisting of $N$ legitimate UAV users, a destination ground base station, a friendly UAV jammer, and a malicious ground eavesdropper. These legitimate UAV users communicate with the destination base station through $N$ orthogonal channels while facing the risk of eavesdropping by the eavesdropper. The friendly UAV jammer has the capability to transmit cooperative jamming signals simultaneously over these channels to enhance the overall secrecy rate against the eavesdropper. The optimization objective is to maximize the aggregate security rate of all legitimate users through optimal power allocation of the friendly jammer across each channel.

\par \textbf{Framework Operation Flow:} The LLMs interactive module in Fig. \ref{fig_4} shows the optimization flow of the proposed framework to solve the UAV secure communication optimization problem. In traditional optimization process, a network designer needs to formulate a mathematical optimization problem, use an appropriate game theory model to analyze the problem, and design effective algorithms to solve it, which relies on rich professional knowledge. With the help of the proposed framework, the optimization process can be automatically implemented through three rounds of interaction between the network designer and the GAI agent. Therefore, the proposed framework can effectively solve the challenge faced by the traditional optimization process. 

\par \textbf{Simulation Setup:} To validate the effectiveness of the proposed framework, we conducted relevant simulation experiments. The specific simulation setup are described as follows.
\begin{itemize}
    \item \textit{Framework Setup}: For the proposed framework, the pluggable LLM module is implemented  by OpenAI APIs for calling the ChatGPT-4 model, and the RAG module is built based on LangChain. 
    \item \textit{Key Parameters}: The size of chunks in the RAG is set to 2000, and the GAI agent retrieves 3 chunks to perform inference in each round. The total power of the friendly UAV jammer is set as $P_{\text{total}} = 0.05\ \text{W}$, and the transmission power of the legitimate UAV users is set as $P_s = 0.01\ \text{W}$. Moreover, the channel coefficients from the legitimate UAV users to the destination base station, from the legitimate UAV users to the eavesdropper and from the UAV jammer to the eavesdropper are set as $g_s, g_e, g_{e,i} \in  [0, 1]\times 10^{10}\ \text{W}$.
    \item \textit{Evaluation Metrics}: To validate the effectiveness of the proposed framework, we evaluated the following performance metrics. \textit{1) Convergence:} Convergence evaluates whether  the algorithm generated by the proposed framework can achieve a Nash equilibrium solution. \textit{2) System Sum Secrecy Rate:} The system sum secrecy rate represents the optimization objective of the considered scenario, with a higher secrecy rate indicating better performance.
\end{itemize}

\par \textbf{Evaluation Results:} The evaluation result module (a) of Fig. \ref{fig_4} shows the convergence of the algorithm generated by the proposed framework (i.e., GAI algorithm). From this figure, we can see that the GAI algorithm reaches a stable state, i.e., Nash equilibrium, after five iterations. The simulation results illustrate that the proposed framework can obtain a Nash equilibrium solution to the game theory problem. 

\par Moreover, the evaluation result module (b) of Fig. \ref{fig_4} shows the performance of the GAI approach with respect to different number of channels, where the expert approach indicates that the algorithm designed by the network experts can be regarded as the optimal approach, the EPR approach indicates that the average power allocation strategy is a benchmark approach, and the multi-agent deep deterministic policy gradient (MADDPG)-based approach (i.e., the MADDPG-based approach) is a deep reinforcement learning method, which is used as a comparative approach. Correspondingly, we can observe that as the number of channels increases, the system sum secrecy rate shows an upward trend. This is because more legitimate UAV users participate in communication. Further, the GAI-based approach has approximately the same performance as the expert algorithm. This can be attributed to the fact that the proposed framework can generate accurate solution approaches for the considered scenario by learning from a large amount of relevant information.

\par The GAI-based approach outperforms the MADDPG-based approach and the EPR approach, and the reasons are as follows. First, due to the heterogeneous communication conditions of legitimate UAV users, the average power allocation strategy of EPR struggles to efficiently utilize the limited power resources. Second, the MADDPG-based approach requires agents to continuously interact and learn to discover the optimal strategy. This dynamic learning process makes it difficult for the agents to achieve effective strategy coordination. In conclusion, the evaluation results demonstrate the effectiveness of the proposed framework.

%
%

\subsection{Discussion}

\par In this section, we further discuss the proposed framework including the requirements and new challenges as well as the associated costs.

\par \textit{1) Requirements and Challenges}: The proposed framework introduces the following requirements and novel challenges:

\begin{itemize}
	
    \item \textit{Accurate and Diverse Database}: In the proposed framework, the LLMs need to be able to handle and understand diverse optimization problems and game theory models. Therefore, the accuracy and diversity of the database provided by RAG are crucial to ensure that the generated content is not only correct but also covers a wide range of application scenarios. However, obtaining accurate and diverse database is challenging.
    
    \item \textit{Detailed and Precise Requests}: The proposed framework necessitates interaction with the user to generate appropriate responses. Therefore, satisfactory generation outcomes require the user to provide detailed and precise requests. However, ensuring that the users can provide sufficiently detailed and precise requests is challenging.
    
    \item \textit{Additional Parameter Tuning}: The integration of LLMs and RAG models introduces additional parameters, such as the size of knowledge chunks and the number of retrieval blocks. These parameters not only influence the response time but also affect the accuracy and consistency of the LLMs. Therefore, achieving optimal performance requires careful parameter tuning tailored to specific optimization problems, which is also challenging.
    
\end{itemize}

\par \textit{2) Associated Costs}: Implementing the proposed framework mainly involves the following costs:

\begin{itemize}
	
    \item \textit{Database Collection Costs:} RAG necessitates maintaining a large-scale database to store extensive knowledge related to network optimization and game theory, including academic papers, technical documents, and algorithms, in order to enhance the accuracy of LLM-generated responses. Therefore, the construction of the database, including information collection and filtering, incurs significant costs.
    
    \item \textit{Computational Costs:} RAG extracts knowledge which is relevant to user requests from the large-scale database, and the LLMs generate responses to these requests. This process often requires substantial computational resources, leading to high computational costs.
    
\end{itemize}

%
%

\section{Future Directions}

\par In this section, we present three major future directions for the enhancement and extension of GAI-enabled game theory.

%
%

\subsection{GDM-enhanced Non-Cooperative Game}

\par For the non-cooperative game for the networking scenarios with numerous players and complex decision spaces, obtaining a Nash equialibrium is often computationally challenging due to the need to search through a high-dimensional and exponentially large space. GDM has outstanding capabilities in modeling complex data distribution and data generation~\cite{cao2024survey}. With the help of GDM, it is possible to enhance the search efficiency for Nash equilibrium. Therefore, future work could focus on incorporating GDM into the solution process of non-cooperative games, aiming to identify solutions that satisfy Nash equilibrium conditions through sampling and optimization of the strategy space.

%
%

\subsection{Performance Evaluation of GAI-enabled Game Theory}

\par It is crucial to design a comprehensive performance evaluation model to evaluate the performance of GAI combined with game theory. This is because effective evaluation models can provide directions for guiding and improving the GAI-enabled game theory framework. However, traditional approaches often rely on human expertise to formulate evaluation models, which tend to lack flexibility. Therefore, future work could explore the integration of human feedback with existing evaluation models by utilizing LLMs and RAG technology to design more intelligent evaluation models.

%
%

\subsection{GAI-enabled Game Theory for Future Networks}

\par As a key technology for future 6G wireless communications, the space-air-ground integrated network (SAGIN) has been proposed to provide seamless communication coverage and high-speed connections~\cite{liu2018space}. However, due to the heterogeneous network components and the dynamic network environment, optimizing SAGIN is a challenging task. Therefore, future research could explore the use of GAI-enabled game theory to model and optimize network scheduling, routing strategies, and resource allocation within the dynamic network environment of SAGIN.

%
%

\section{Conclusion}

\par In this article, we explored the integration of LLMs with game theory, and concluded the advantages of LLMs for game theory. Following this, we presented the potential applications of LLMs combined with game theory in networking. Based on these, we proposed a framework that utilized RAG to combine LLMs with game theory, aiming to achieve the intelligence of game theory in network applications. We conducted a case study on UAV secure communication optimization to validate the effectiveness of the proposed framework. Finally, we discussed several potential research directions for future extensions.

\bibliographystyle{IEEEtran}
\bibliography{ref}

%
%

\section*{Biographies}

\noindent 
\textsc{Long He} (\text{helong0517@foxmail.com}) received a BS degree in Computer Science and Technology from Chengdu University of Technology, Sichuan, China, in 2019. He is currently working toward the PhD degree in Computer Science and Technology at Jilin University, Changchun, China. His research interests include vehicular networks and edge computing.

\vspace{1em}

\noindent 
\textsc{Geng Sun} [S'17, M'19, SM'24] (\text{sungeng@jlu.edu.cn}) received a B.S. degree in Communication Engineering from Dalian Polytechnic University, China, and the Ph.D. degree in Computer Science from Jilin University, China, in 2011 and 2018, respectively. He was a visiting researcher in the School of Electrical and Computer Engineering at Georgia Institute of Technology, USA. He is currently a Professor in the College of Computer Science and Technology at Jilin University, and his research interests include UAV Networks, collaborative beamforming, generative AI and optimizations.

\vspace{1em}

\noindent 
\textsc{Dusit Niyato} [M'09, SM'15, F'17] (\text{dniyato@ntu.edu.sg}) is currently a professor in the School of Computer Science and Engineering, Nanyang Technological University, Singapore. He received the B.Eng. degree from King Mongkut's Institute of Technology Ladkrabang (KMITL), Thailand, in 1999, and the Ph.D. in electrical and computer engineering from the University of Manitoba, Canada, in 2008. His research interests are in the areas of Internet of Things (IoT), machine learning, and incentive mechanism design.

\vspace{1em}

\noindent 
\textsc{Hongyang Du} (\text{duhy@eee.hku.hk}) is an assistant professor at the Department of Electrical and Electronic Engineering, the University of Hong Kong. He received his Ph.D. degree from the College of Computing and Data Science, Energy Research Institute @ NTU, Nanyang Technological University, Singapore, in 2024. He serves as the Editor-in-Chief assistant of \textit{IEEE Communications Surveys \& Tutorials} (2022-2024), and the Guest Editor for \textit{IEEE Vehicular Technology Magazine}. His research interests include edge intelligence, generative AI, and network management.

\vspace{1em}

\noindent 
\textsc{Fang Mei} (\text{meifang@jlu.edu.cn}) received the M.Sc. and Ph.D. degrees in computer science from Jilin University, Changchun, China, in 2005 and 2010, respectively. She is currently an Associate Professor in the College of Computer Science and Technology at Jilin University. Her research interests include intelligent information processing, multi-access edge computing, vehicular networks, and antenna arrays.

\vspace{1em}

\noindent 
\textsc{Jiawen Kang} (\text{kavinkang@gdut.edu.cn}) received the Ph.D. degree from the Guangdong University of Technology, China, in 2018. He has been a postdoc at Nanyang Technological University, Singapore, from 2018 to 2021. He is currently a full professor at Guangdong University of Technology, China. His research interests focus on blockchain, security, and privacy protection.

\vspace{1em}

\noindent 
\textsc{Mérouane Debbah} [S'01, M'04, SM'08, F'15] (\text{merouane.debbah@ku.ac.ae}) received the M.Sc. and Ph.D. degrees from the Ecole Normale Supérieure Paris-Saclay, France. Since 2023, he is a Professor at Khalifa University of Science and Technology in Abu Dhabi and founding director of the 6G center. He has managed 8 EU projects and more than 24 national and international projects. His research interests lie in fundamental mathematics, algorithms, statistics, information, and communication sciences research. He holds more than 50 patents. He is an IEEE Fellow, a WWRF Fellow, a Eurasip Fellow, an AAIA Fellow, an Institut Louis Bachelier Fellow and a Membre émérite SEE.

\vspace{1em}

\noindent 
\textsc{Zhu Han} [S'01, M'04, SM'09, F'14] (\text{zhuhan22@gmail.com}) currently is a professor in the Electrical and Computer Engineering Department at the University of Houston, Texas. He has been an AAAS Fellow since 2019. He received the IEEE Kiyo Tomiyasu Award in 2020. He has been a 1 percent highly cited researcher since 2017 according to Web of Science.
\end{document}